\newcommand \be{\begin{equation}}
\newcommand \ba{\begin{eqnarray}}
\newcommand \ea{\end{eqnarray}}
\newcommand \ee{\end{equation}}

 \documentclass[ final,5p,times,twocolumn ]{elsarticle}
\usepackage{float}
\usepackage{color}
\usepackage{epsfig}
\usepackage{url} 
\usepackage{amsmath} 
\begin{document}
\begin{frontmatter}
\title{Edge excitation geometry for studying intrinsic emission spectra of bulk {\em n}-InP}
\author{Oleg Semyonov}
\author[cor1]{Arsen Subashiev}
\fntext[cor1]{Corresponding author, arsen.subashiev@stonybrook.edu}
\author{Zhichao Chen}
\author{Serge Luryi}

\address{Department of Electrical and Computer Engineering,
State University of New York at Stony Brook, Stony Brook, NY,
11794-2350}
\begin{abstract}
The shape of the photoluminescence line excited at an edge face of InP wafer and registered from the broadside is used to investigate the intrinsic emission spectrum. The procedure is  much less sensitive to the surface properties and the carrier kinetics than the conventional methods used with the reflection or transmission geometry of photoluminescence.  Our method provides a tool for studying the effects of non-equilibrium distribution of minority carriers in doped direct-band semiconductors.
\end{abstract}
\end{frontmatter} 

\section*{Introduction}
Luminescence spectroscopy of optically excited semiconductors is widely used for characterization of semiconductor materials \cite{Pavesi}, as well as for studying effects  associated with  the impurity levels \cite{Zhang}, doping \cite{Haufe}, minority carrier kinetics \cite{Hwang}, 
and level quantization in confined structures \cite{Gershoni}. All these studies rely on the possibility to determine the intrinsic emission spectrum produced in radiative recombination. 

For a quasi-equilibrium carrier energy distribution, the intrinsic emission spectra are closely related to the absorption spectrum through the so-called van Roosbroek-Shockley  (VRS) relation, also referred to as the Kennard-Stepanov rule (mostly in molecular physics) and as the Kubo-Martin-Schwinger theorem 
(reflecting a more fundamental underlying principle) \cite{VRSh}. It should be noted that this relation does not apply when the optically excited system is significantly away from equilibrium, e.g., when the energy relaxation and/or spatial equilibration of carriers is slow compared to recombination. 

Even when the quasi-equilibrium of carrier ensembles is reasonably well established, in bulk semiconductors there remains the problem that the absorption spectrum is rarely known in the entire energy range from  the interband region, where the absorption coefficient is very high, to the deep tailing region with exponentially small absorption. Available theories (e.g. \cite{Gauss,Lax,SJohn,Greef}) are of little help as they do not provide description of the absorption in the entire energy range. For thin  (compared to the absorption length) layers this difficulty can be successfully overcome \cite{Pfeiffer} using the excitation spectra (the dependence of luminescence intensity on the excitation wavelength) as a measure of the absorption spectrum. Such a luxury is not available for thicker layers, both because of the non-uniform excitation of minority carriers and the spectral filtering of the outgoing radiation via self-absorption. 

For thin samples there is fortunately little filtering and the observed luminescence spectra of epitaxially grown submicron or quantum-well layers are fairly close to the intrinsic emission spectra (see Ref. \cite{Sieg} for a discussion) and their interpretation is unambiguous. 

The possibility of measuring both the absorption and the emission spectra in thin layers enables an experimental check of the VRS relation in the entire energy range of luminescence. Detailed studies of QW emission showed that the relation holds with high accuracy in modulation-doped samples \cite{Pfeiffer} where the quasi-equilibrium emission conditions are fulfilled. 

On the other hand, the VRS relation is violated in homogeneously doped QW structures at low temperatures \cite{khitrova}, apparently due to spatial fluctuations of the potential leading to a deviation from equilibrium of the spatial distribution of carriers. For these structures, when the majority carriers are highly degenerate, one obtains a better fit to the experimental spectra of thin doped epitaxial layers by using an emission model \cite{monemar} that totally neglects the momentum conservation in optical transitions. In this case, clearly, the emission spectrum does not obey VRS. A more accurate description of the spectrum in the tailing region can be obtained taking into account the effects of Coulomb interaction between majority carriers \cite{Haufe}. This also justifies the model \cite{monemar}.
   
In bulk wafers both the intensity and the spectra of emission are often influenced by unknown defects of poorly controlled concentration. This makes the intrinsic emission spectra even less predictable. Interpretation of the experimentally observed spectra is also less straightforward as they are modified by spectral filtering due to the wavelength-dependent reabsorption of the outgoing radiation \cite{Bebb,vonRoos,Semyon1}. The reabsorption process strongly suppresses the blue wing of the emission spectrum (where the absorption is high) and results in a noticeable red shift of the emission maximum. Therefore, the spectra are sensitive to the distribution of minority carriers excited in the sample, which in turn depends on the carrier kinetics and surface conditions. Correct account of the minority-carrier transport becomes essential.

For normal diffusion of holes, the distribution is exponential and is determined by the hole generation profile, the diffusion length (typically on a micron scale), and the surface recombination velocity.  The spatial distribution of minority carriers varies with the optical excitation energy leading to an excitation energy dependence of the luminescence spectra. Therefore, an unambiguous interpretation of these spectra requires an accurate consideration of the excitation conditions, the carrier transport and the escape mode of the luminescence radiation, providing a set of parameters for fitting the spectra. Usually, these parameters are not well known. 

Fortunately, in crystals with high quantum efficiency of emission we find a way of avoiding this difficulty, based on the giant spread of the minority carriers in the sample. The minority carrier distribution is strongly modified by the recycling effects (multiple emission-reabsorption events) \cite{Asbeck,Rossin,Luryi-IJHES}. When the losses of carriers (via non-radiative recombination) and photons (via residual free-carrier absorption) are low, the recycling leads to anomalous diffusion and a dramatic increase of the minority-carrier spread in the sample (up to centimeter-scale distances) \cite{Semyon1}.  This was found to be crucial for the correct interpretation of the difference between the observed spectra in the transmission and reflection geometries \cite{Luryi-IJHES}. 

The much enhanced spread of minority carriers makes it possible to study the luminescence spectra in a novel geometry, where the excitation is provided at an edge face of the sample and observation is carried out from the broadside \cite{LuryiPRB,AVS}. The observed spectra do not depend on the distance from the edge to the observation region in a wide range of distances, for differently doped samples and at different temperatures, which implies a stable minority carrier distribution \cite{LuryiPRB,AVS}. 

Here we present a study of the luminescence emission spectra for a moderately  doped $n$-InP wafer, obtained with the edge-face excitation. Since the outgoing radiation is captured from the crystal region that is remote from the edge, the effective excitation in that region is predominantly by the radiation in the red wing of the emission spectrum, for which the wafer is mostly transparent. Therefore, the excitation spreads to a high extent homogeneously across the wafer. The main advantage of this geometry is a much simplified and more accurate account of the spectral filtering.  

We report the temperature dependence of the spectra. At temperatures $T>200$ K the spectra are close to those obtained with van Roosbroek-Shockley relation for the intrinsic emission spectrum. However for lower temperatures, the deviation is noticeable. We interpret this as due to non-equilibrium effects arising from the spatial fluctuations of the impurity potential.  
\begin{figure}[t,b]
\epsfig{figure=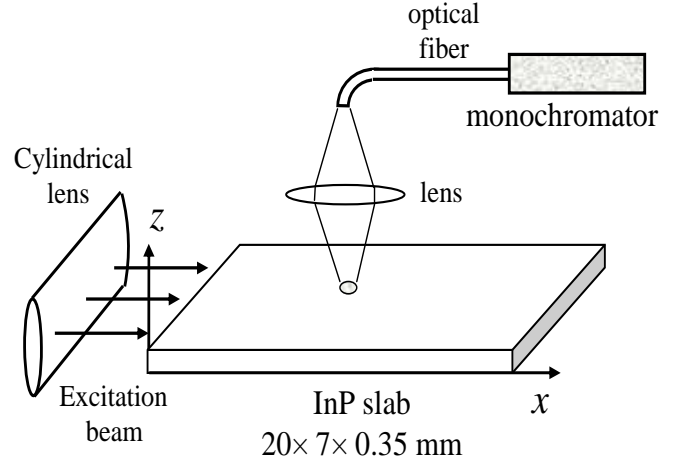,width=8.9cm,height=6.2cm} 
\caption[]{(Color online) Schematics of the experimental geometry. An 808 nm excitation beam is focused on a 0.35$\times$7 mm$^2$ edge of the sample; the photoluminescence is registered from the broadside at varying $x$.} 
\label{geom}
\end{figure}
\section*{Experiment}

We studied moderately-doped single-crystal $n$-InP wafers  \cite{NIKKO}, doped with sulfur to $N_{\rm D}=3\times 10^{17}$ cm$^{-3}$. The room-temperature Hall mobility of 3000 cm$^2$/Vs is comparable to those previously reported for InP epitaxial films with similar doping concentrations \cite{Sieg}. 

The edge-excitation geometry illustrated in Fig. \ref{geom} allowed us to study luminescence far away from the excitation region. The broadside luminescence was captured by a lens to image a small patch of the surface onto an optical fiber transmitting the light to the entrance slit of a monochromator. After passing through the output slit, the chopped and spectrally dispersed light beam illuminated a 1.1-cm Si-detector, connected to a lock-in amplifier. To account for the spectrally-dependent attenuation in the optical system and the spectral sensitivity of the detector, we scanned the spectrum of a halogen lamp, which in the spectral range 800 to 1100 nm is quite close to the spectrum of black-body radiation at $T=2500$ K. The obtained spectral response of the whole system was used to correct the raw spectral shapes and reconstruct the true luminescence spectra.

The absorption spectra and the luminescence spectra measured in a more traditional reflection geometry were also available for the same sample \cite{Semyon1,Semyon2,AVSUrb}. We remark that although the shapes of both absorption and luminescence spectra at room temperature had a negligible variation in the doping interval $N_{\rm D}=2$ to $6\times 10^{17}$ cm$^{-3}$, the samples with $N_{\rm D}=3\times 10^{17}$ cm$^{-3}$  had the brightest emission and highest quantum efficiency \cite{HeavyDoping}. 

\begin{figure}[t]
\epsfig{figure=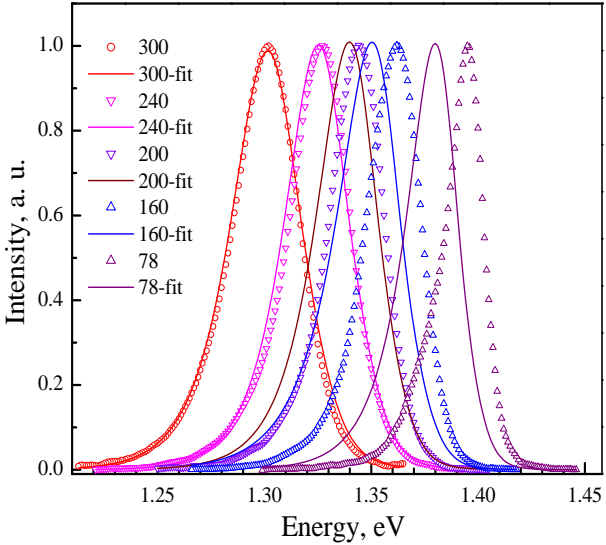,width=8.cm,height=7.3cm} 
\caption[]{(Color online) Edge excitation spectra at different temperatures (dots). Also shown (by solid lines) are results of calculations based on the intrinsic VRS spectrum (\ref{VRS}).} 
\label{SpecAllT}
\end{figure}

Luminescence spectra for $N_{\rm D}=3\times 10^{17}$ cm$^{-3}$ sample at various temperatures, excited by a 808-nm laser beam and measured at $x=0.5$ mm, are shown in  Fig. \ref{SpecAllT}.  Also shown are the results of calculations discussed below. The spectra recorded at different distances $x$ between the observation spot and the excitation edge in the range between $x=0.2$ and $5$ mm were all identical in shape. No shape variation with the laser excitation energy $E_{\rm ex}$ (for the wavelengths $\lambda=$ 808, 780, 670 and 640 nm) was observed either.

As to the variation of luminescence intensity with the excitation energy, it cannot be used to deduce the absorption spectrum, in contrast to the case of thin films \cite{Pfeiffer}. The dependence arises mainly because of the excitation-dependent competition between the surface and bulk channels of recombination \cite{total}. 
The blue shift of the luminescence line clearly observed at lower temperatures is associated with the increasing band-gap. Also observed but less pronounced is some narrowing of the line. The red wings of the lines at all temperatures closely follow an exponential decay law 
\be
I\propto \exp [(E-E_g)/{\Delta'(T)}]~. \label{RedWing}
\ee
Similar exponential law with the same parameter $\Delta'(T)$ was observed in reflection-geometry experiments \cite{Semyon1}.

\section*{Discussion: Effects of spectral filtering}
Because of the effects of self-absorption and multiple surface reflection, the luminescence spectra $I(E)$ registered from the broadside of a slab of finite thickness $d$ are modified compared to the intrinsic emission spectra $S_i(E)$. The relation is of the form \cite{Semyon1}:
\be
I(E)= S_i(E)F(E)~. \label{SpecR} 
\ee
The function $F(E)$ describes the spectral filtering. It depends on the details of excitation and the observation geometry. Due to the high refractive index $n\approx 3.4$ of InP, the escape cone of luminescent radiation is narrow, so that only the fraction of radiation, which propagates nearly perpendicular to the surface, goes out. The filtering function $F(E)$ can be expressed through a one-pass filtering function $F_1(E)$ for the light that comes to the surface without reflections, 
\ba 
F_1(E)&=& \int_0^d p(z)\exp[-\alpha(E)z] dz, \label{Filt1}
\ea
where $p(z)$ is the photo-generated hole concentration at a distance $z$ from the broadside surface. In our case, the distribution $p(z)$ is almost homogeneous along $z$ and symmetric relative to the mid-plane $z=d/2$. Taking into account multiple reflections of the luminescence radiation, we have \cite {Semyon1}
\be  
F(E)=F_1~\frac{1-R(E)}{1-R(E) \exp(-\alpha d)}~, \label{refFil}
\ee
where $R(E)$ is the surface reflection coefficient (confidently assumed to be the same for both sample surfaces).    

At large distances $x$ from the excitation edge, the distribution $p(z)$ that appears in Eq. (\ref{Filt1}) is generated by weak reabsorption of red photons in the transparency region. Therefore, $p(z)$ is nearly uniform. It deviates from a constant only due to surface recombination and only in a narrow depletion region near each broadside surface,
\be  
p(z)=p_0[1-c \exp(-z/l)]~, \label{DifDist}
\ee
where $l$ is the width of the region with depleted minority carrier concentration. This width is analogous to the conventional diffusion length. If the surface recombination velocity $s$ is fast, the constant $c$ can be estimated as $c=sl/(sl+D) \le 1$, where $D$ is the diffusion coefficient of holes. Both $c$ and $l$ are used in this work as (non-essential) fitting parameters. 

Evidently, surface recombination effects will be noticeable in our geometry only in the spectral region where $\alpha(E) l \ge 1$ (i.e. in the far end of the blue wing of the observed line). In contrast, in the traditional reflection-geometry experiment this effect modifies the luminescence line along the whole blue wing starting from the line center. In that geometry, the minority-carrier concentration rapidly decays away from the surface and the spectra are very sensitive to the shape of $p(z)$. That shape is defined by the kinetics of surface recombination and diffusion, as well as by specific conditions of the excitation. 
\begin{figure}[t]
\epsfig{figure=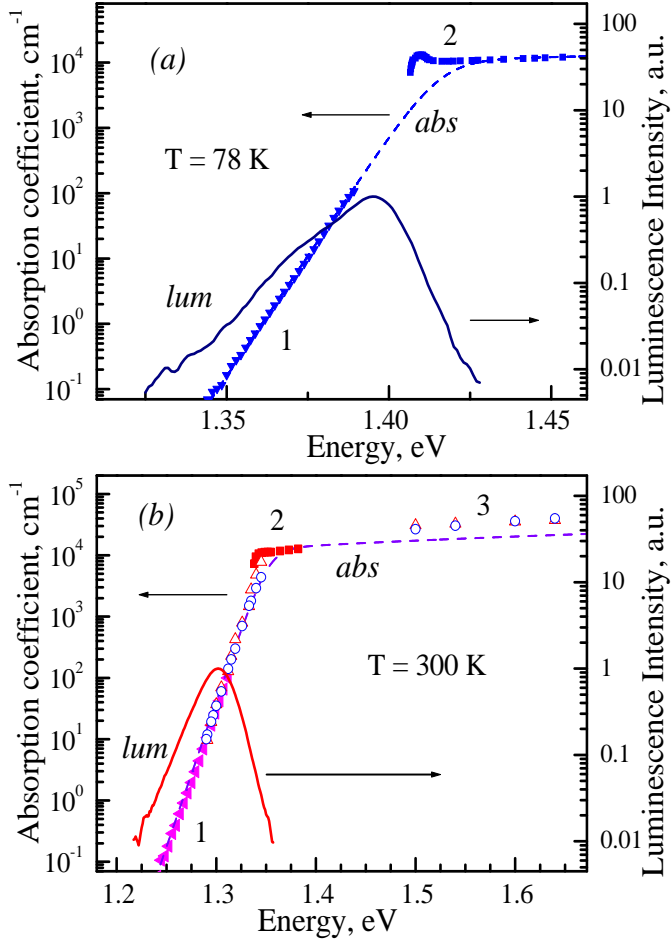,width=8.8cm,height=12.5cm} 
\caption[]{(Color online) Comparison of the luminescence  spectra observed ($a$) at $T=300$ K and ($b$) at $T=78$ K with the corresponding absorption spectra  for $n$-InP sample with $N_{\rm D}=3 \times 10^{17}$  cm$^{-3}$. Shown are the luminescence spectra [log scale, curves marked as $lum$] measured in the edge-excitation geometry, the experimental absorption spectra from \cite{AVSUrb} [triangles, curves (1)],  as well as the absorption spectra for undoped InP, taken (2) from \cite{Turner} and (3) from \cite{TB}. Dashed lines (marked as $abs$) show the interpolated absorption coefficient used in calculations of the luminescence spectra and the filtering functions.}
\label{Spec78vs300}
\end{figure}

Equation (\ref{SpecR}) suggests that the intrinsic emission spectrum can be recovered from the observed luminescence spectrum, so long as the filtering function is calculated with a sufficient accuracy.  Let us now discuss the limitations of this procedure. 

The main limitation comes from the steep growth of the absorption coefficient in the range of interest, leading to a steep variation of $F(E)$. The interband absorption spectra of direct-bandgap semiconductors typically feature an exponential energy dependence (``Urbach tail'') \cite{Urbach} extending at least from $\alpha = 1$ to 500 cm$^{-1}$. In our moderately doped InP samples, this Urbach tail is longer in both directions, extending up to 5000 cm$^{-1}$ for all $T$.  One can evaluate the interband absorption coefficient deeper into the bandgap by subtracting the residual (free-carrier) absorption that is essentially constant in this energy region and grows linearly with the doping \cite{Semyon2}. 

The interband absorption spectra for our sample are shown in Fig. \ref{Spec78vs300} for $T$=300 K and $T=78$ K. For comparison and further discussion, we also show the absorption coefficient in the region of large absorption for a much lower-doped sample ($N_{\rm D}=5\times 10^{15}$ cm$^{-1}$) from \cite{Turner} exhibiting an excitonic feature near the absorption edge, as well as the relevant textbook data \cite{TB} (where this feature is not resolved).  

Matching the experimental absorption to the Urbach law, 
\be \alpha (E) = \alpha_0 \exp \left[ \frac{E-E_g(n,T)}{\Delta(n,T)}\right], \label{UrbTail} 
\ee 
where $\Delta(n,T)$ is the Urbach tail parameter, with the well-known behavior of the bandgap energy $E_g(T)$ for undoped InP \cite{Vurgaft} gives  $\alpha_0=1.1\times 10^4$ cm$^{-1}$; this value provides a good fit in a wide temperature range $T = 0$ to 1000 K \cite{Beaudoin}. The physical interpretation of $\alpha_0$ in Eq. (\ref{UrbTail}) as the value of $\alpha$ at $E=E_g$  (i.e. above the steep slope region) suggests that $\alpha_0$ should not vary with the concentration at a moderate doping level,  $n \le 10^{18}$ cm$^{-3}$, when the Fermi level is still below or near the conduction band edge. This allows us to further estimate both $E_g(n,T)$ and $\Delta(n,T)$ from the absorption spectra of our moderately $n$-doped sample. The results are shown in Fig. \ref{DelEg} together with the data obtained for semi-insulating InP  \cite{Beaudoin}. One can see that in the temperature interval $300\ge T \ge 150 $ K the values of $\Delta(n,T)$ are increased by $\approx$ 1.5 meV as compared to the undoped sample, while  the difference in $E_g(T,n)$ is minimal. Variations of both $E_g(n,T)$ and $\Delta(n,T)$ with the temperature and the doping can be described \cite{Beaudoin,Chung} in  the Einstein model for lattice vibrations,
\begin{equation}
\begin{aligned}
\refstepcounter{equation}
\label{varNT}
E_g(n,T)&  =  E_{g,0}-\Delta E_g(n,T)- s_g k\theta \left[ \coth (\theta/2T)-1\right],
\nonumber \\ 
\Delta(n,T)& =  s(n) + s_\Delta k \theta \left[ \coth(\theta/2T)-1\right]~,\hspace{2cm} (7)
\end{aligned}
\end{equation}
where $\theta$ is an effective Debye temperature, which for our samples equals $270$ K \cite{Debye}. The term $\Delta E_g(n,T)$ accounts for bandgap narrowing with the doping. Terms with $s_g$ and $s_\Delta$ describe the effects of temperature and are proportional to the electron-phonon coupling constant. Term with $s(n)$ describes the effect of concentration on the Urbach tail parameter. Equations (\ref{varNT}) imply that these two effects are statistically independent and can be included additively \cite{SJohn}. The observed reduction of $s (n)$ at low temperatures is related to the reduced screening length of the random potential.

Based on the absorption spectrum, we note that for $E \ge E_g$ the region $z\ge 1/\alpha_0$ gives an exponentially small contribution to the integral in Eq. (\ref{Filt1}), while the experimentally known values of $\alpha(E)$ are not reliable. Therefore, in an attempt to recover the intrinsic spectrum from Eq. (\ref{SpecR}), the blue wing of the intrinsic emission spectrum would not be accurately estimated. It is more consistent to perform a model calculation of the intrinsic spectrum and then compare with the experiment using Eq. (\ref{SpecR}). While this procedure is accurate, it gives no prejudice about the suppressed blue wing of the intrinsic spectrum.

\section*{Deviation from  thermal equilibrium}
For a quasi-equilibrium carrier energy distribution (with different quasi-Fermi levels for electrons and holes) the intrinsic emission spectrum obeys the VRS relation \cite{VRSh},
\be
S_i(E) = \alpha(E)E^3 \exp(-E/kT),
\label{VRS}
\ee 
which allows one to calculate the spectral shape of intrinsic emission using experimental data for the absorption spectrum.
\begin{figure}[t,b]
\epsfig{figure=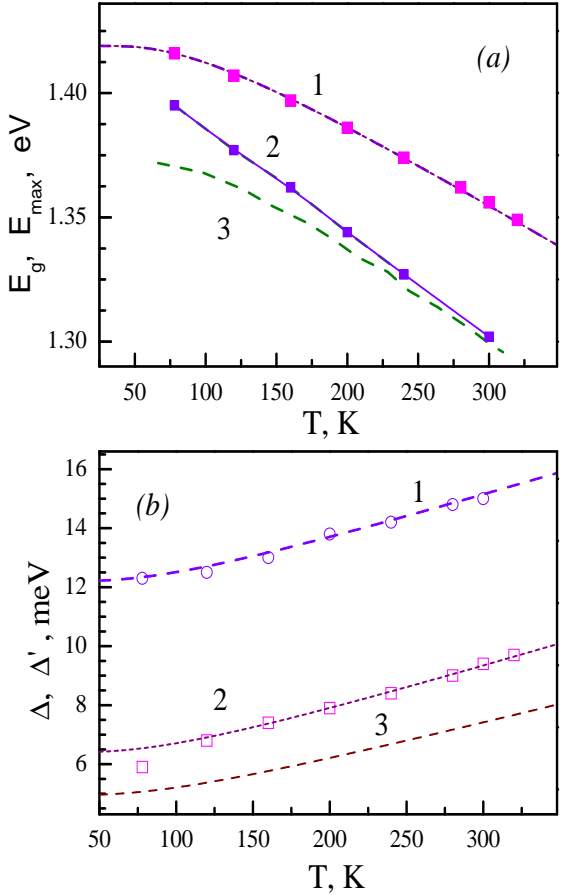,width=8.6cm,height=12.5cm} 
\caption[]{(Color online) Temperature variation of the parameters of luminescence spectra for $n$-type InP sample, $N_{\rm D}=3\times10^{17}$ cm$^{-3}$. Curves (1) and (2) in panel $(a)$ show, respectively, the bandgap $E_{g}$ obtained from the absorption spectra and the position $E_{\rm max}$ of the line maximum; dashed curve (3) indicates the $E_{\rm max}$ expected for a homogeneous impurity distribution. Panel $(b)$ shows (curve 1) the red-wing exponential decay parameter $\Delta'$ and (curve 2) the Urbach tail parameter $\Delta$. Experimental data are indicated by the symbols while the dashed lines represent fitting by Eqs. (\ref{varNT}) with $\theta=270$ K. Curve 3 shows the Urbach tail parameter $\Delta$ for undoped InP, from Ref. \cite{Beaudoin}. } 
\label{DelEg}
\end{figure}
The calculated edge-excited luminescence spectra are presented in Fig. \ref{SpecAllT} by the solid lines. The calculation is based on the interpolated absorption coefficient \cite{Semyon1}, shown in Fig. \ref{Spec78vs300} for $T =$78 K and 300 K. For not too low temperatures ($T \ge 200$ K), the spectra calculated from the intrinsic VRS spectra (\ref{VRS}) provide an excellent fit to the shape of the observed line. The only fitting  parameter is the surface depletion length $l$.  Because of the recycling effects, one can anticipate somewhat enlarged values of $l$, compared to estimates based on typical surface recombination velocity and normal diffusion coefficient of holes, so that the fitting values in the range of 7 to 14 $\mu$m appear reasonable. We shall not be concerned with the more detailed interpretation of the parameter $l$ because its influence is limited only to the far blue wing of the line and the experimental accuracy  of the spectrum  measurement (\ref{SpecR}) is insufficient in that region. The position of the line center can be calculated as in Ref. \cite{LuryiPRB}, see the dashed line (3) in Fig. \ref{DelEg}(a). The line center energy is approximately described by $\alpha(E)d=1$, which again shows that it is controlled by the filtering. 

For $T\le$ 200 K the calculated lines no longer agree with the experiment. The experimental line maxima (Fig. \ref{DelEg}) are systematically at higher energies than those calculated from the VRS relation (\ref{VRS}). At 78 K this discrepancy is about 13 meV. 

Moreover, the relation between parameters $\Delta'$ and $\Delta$ expected \cite{AVSUrb} from the VRS relation no longer holds at low temperatures. This relation is of the form
\be
\frac{1}{\Delta'(n,T)}=\frac{1}{\Delta(n,T)}-\frac{1}{T}~.
\label{DelPri}
\ee
We find that for $T\le$ 200 K the values of $\Delta'(T)$ extracted from the spectra are smaller than those calculated from the red-wing slope of log $\alpha$ with Eq. (\ref{DelPri}).  This may suggest that holes have an effective temperature higher than the lattice temperature. However, the energy relaxation time $\tau_\epsilon$ of holes at room temperature is several orders of magnitude shorter than the radiative recombination time $\tau_{\rm rad}$. The ratio $\tau_{\rm rad}/\tau_\epsilon$ is reduced at lower temperatures, since $\tau_{\rm rad}$ gets shorter while the energy relaxation slows down.  Still,  the ``hot hole'' stationary state seems to be highly improbable, even though the decreasing ratio $\tau_{\rm rad}/\tau_\epsilon$ can have an effect of reducing the spectral filtering. Such a reduction was experimentally observed \cite{Moor} in the standard reflection geometry for low-temperature (78 K and below) emission from InSb. Quantitative account of this effect requires a separate study, but qualitatively it would explain the blue shift of the experimental line maxima at low temperatures compared to their position calculated using Eq. (\ref{Filt1}).

Accounting for a discrepancy with the VRS relation requires a model of the non-equilibrium state. As already mentioned, we consider unlikely a homogeneous ensemble of holes characterized by an elevated effective temperature, because that would require a physically unrealistic small ratio $\tau_{\rm rad}/\tau_\epsilon \approx 1$. 

A more likely candidate for a non-equilibrium stationary state of holes at low temperatures is associated with the spatial fluctuations \cite{fluct,elisseev,khitrova} in doping concentration and the resultant local potential fluctuations. The ionized donors repel the holes thus increasing their potential energy. A local depletion of the donor concentration provides a potential well from which holes recombine. The recombination process is faster than that of spatial exchange of holes between different wells and therefore the VRS relation need not be obeyed. 

Moreover, since the lower doping results in lower values of the Urbach parameter $\Delta(n,T)$ in absorption spectra, the local decrease of doping should lead to smaller values of $\Delta'(T)$. One can use this observation to check whether a local decrease of concentration of ionized donors is sufficient to interpret the observed variation of  $\Delta'(n,T)$. The required  fluctuations of $n$ can be estimated from Eqs. (\ref{DelPri}) and (\ref{varNT}), which incorporate the experimental values of $\Delta'(n,T)$ and $\Delta(n,T)$. For $T$=78 K we find that the local concentration drop by an order of magnitude is sufficient to explain the observed value of $\Delta'(n,T)$ --- without assuming an elevated effective hole temperature.

The local concentration drop can also explain the low-temperature blue shift of the luminescence maxima. The locally low-doped areas of the sample do not experience the bandgap narrowing effect that could be estimated for homogeneous doping \cite{Sieg,Borghs}. Precise values of the narrowing are still a matter of discussion and apparently depend on the experimental conditions for observation. According to optical experiments of Ref. \cite{Sieg,Chung}, the bandgap narrowing in $n$-InP is  well approximated by  $\Delta E_g(n)=- c~(n\times 10^{-18})^{1/3}$ meV. For $T=78$ K the constant $c=71.4$ \cite{Sieg}, giving an estimate of $\Delta E_g\approx 20$ meV for the depleted fluctuation region. We see that the effect is sufficient to account for the blue shift of luminescence lines at $T=78$ K. For $T=300$ K one has $c=22.5$ and the bandgap narrowing effect is reduced \cite{Chung}.

With the increasing temperature, the hole ensemble approaches equilibrium and the spectrum approaches that calculated from Eq. (\ref{VRS}).  

\section*{Conclusions}

We have presented a novel photo-luminescence method for studying the intrinsic spectra in bulk semiconductors. The photoluminescence is excited at an edge face of a semiconductor wafer and registered from the broadside. We show that this geometry has no sensitivity to the excitation energy and little sensitivity to the surface conditions. Moreover, the transverse distribution of minority carriers is well defined. The method is therefore well suited for the characterization of intrinsic emission spectra. It was applied to bulk $n$-type InP wafers of moderate concentration. 

For not too low temperatures, $T\ge 200$ K the measured spectra agree with those calculated on the basis of the van Roosbroek-Shockley relation. For  $T\le 200$ K, the observed spectra deviate from VRS. This deviation is attributed to spatial fluctuations in the doping concentration in our moderately doped InP samples, giving rise to a non-equilibrium distribution of minority carriers. Such fluctuations are unavoidable in a bulk growth of doped crystals; their effect is more pronounced at lower temperatures. They do not cause a deterioration of the quantum efficiency.

\section*{Acknowledgments}
This work was supported by the Defense Threat Reduction Agency through its basic research program and by the New York State Office of Science, Technology and Academic Research through the Center for Advanced Sensor Technology at Stony Brook.


\begin{thebibliography}{99}
\bibitem{HeavyDoping} Heavier doping results in two additional effects that lie outside our present scope. The emission spectra are modified by the conduction-band filling and smearing of the interband absorption edge. The absorption spectra are masked in the tailing region by the residual free-carrier absorption.
\bibitem{total} If we were to record the total intensity of luminescence integrated over the entire broadside of the wafer, the result --- in the absence of surface recombination --- would be independent of the absorption spectrum.
\bibitem{fluct} Spatial fluctuations of the band edges in homogeneously-doped semiconductors have been proposed to explain deviations of low-temperature photoluminescence spectra from the VRS shape both in bulk crystals (with assumed Gaussian absorption tails) \cite{elisseev} and quantum wells \cite{khitrova}.
\end{thebibliography}
\end{document}